\documentclass[%
preprint,
nofootinbib,
 amsmath,amssymb,
 aps,
]{revtex4-1}

\usepackage{graphics}
\usepackage[german, english]{babel}
\usepackage{dsfont}
\usepackage[utf8]{inputenc}
\usepackage{amsmath, amssymb} 	
\usepackage{amsfonts}
\usepackage{braket} 
\usepackage{tensor} 
\usepackage{bbm} 	
\usepackage{graphicx}
\usepackage{color} 
\usepackage{blindtext} 
\usepackage{subcaption} 
\usepackage{comment}

\newcommand{\N}[1]{\mathcal{N}_{#1}(I_{#1},S_{#1})}
\newcommand{\alert}[1]{\textcolor{red}{\textbf{#1}}}
\newcommand{\balert}[1]{\textcolor{blue}{\textbf{#1}}}
\newcommand{\intdd}[1]{\int \hspace{-4pt} \text{d}^{^3} \hspace{-4pt} #1}
 
\renewcommand{\vec}[1]{\boldsymbol{#1}}

\begin{document}

\preprint{}

\title{Applying Twisted Boundary Conditions for Few-body Nuclear Systems  }


\author{Christopher K\"orber, Thomas Luu}
\affiliation{Institute for Advanced Simulations 4, Institute f\"ur Kernphysik 3, Forschungszentrum J\"ulich, D-52425 J\"ulich, Germany }


\date{\today}

\begin{abstract}
We describe and implement twisted boundary conditions for the deuteron and triton systems within finite-volumes using the nuclear lattice EFT formalism.  We investigate the finite-volume dependence of these systems with different twists angles.  We demonstrate how various finite-volume information can be used to improve calculations of binding energies in such a framework.  Our results suggests that with appropriate twisting of boundaries, infinite-volume binding energies can be reliably extracted from calculations using modest volume sizes with cubic length $L\approx8-14$~fm.  Of particular importance is our derivation and numerical verification of three-body analogue of `i-periodic' twist angles that eliminate the leading order finite-volume effects to the three-body binding energy.  
\end{abstract}

\maketitle








\section{Introduction\label{sec:intro}}
Numerical simulations of nuclear observables often utilize finite-volumes (FV) to perform calculations.  Lattice quantum chromodynamics (LQCD) calculations of quarks and gluons, for example, utilize cubic volumes with spatial length $L$ typically of size $\sim 4-6$ fm.  Nuclear lattice effective field theory (NLEFT) calculations using nucleon degrees of freedom, on the other hand, employ volumes that are an order of magnitude larger.  Despite being intrinsically stochastic, both methods have calculated nuclear binding energies of light hadronic systems with impressive, quantitative uncertainties.  Recent LQCD calculations, albeit at unphysical pion masses, have calculated the binding energies of s-shell nuclei and light hyper nuclei \cite{Beane:2012vq,Beane:2012ey,Beane:2010hg,Orginos:2015aya,Etminan:2014tya,Inoue:2011ai,Yamazaki:2015asa,Yamazaki:2012hi}.  Nuclear lattice EFT calculations have readily performed binding-energy calculations of p-shell nuclei \cite{Epelbaum:2009pd,Epelbaum:2011md,Epelbaum:2013paa,Epelbaum:2012iu,Lahde:2015ona} and some medium mass nuclei \cite{Lahde:2013uqa}.  With ever increasing computer resources, calculations of such systems will become even more precise.

All of these calculations, however, suffer from a systematic error that cannot be reduced from increased computer resources:  The calculated energies in a finite volume differ from their infinite volume counterparts.    In principle, this finite-volume (FV) error can be removed by performing calculations of energies in multiple volumes followed by an extrapolation to infinite volume.  In practice this is very difficult due to the large computational costs of performing calculations in multiple volumes.  However, the number of different volume calculations needed to perform a reliable extrapolation may not be exceedingly large if the functional dependence of the FV correction is known.  For the two-body system with periodic boundary conditions, for example, the finite-volume correction to the binding energy is well known and the leading order contribution scales as $\exp(-\kappa L)/L$, where $\kappa$ is the binding momentum \cite{Luscher:1990ux,Beane:2003da}. In \cite{Meissner:2014dea} the functional dependence for three identical bosons in a finite volume (with periodic boundary conditions) and at the unitary limit was derived, and was also determined to fall off exponentially with volume size. For higher A-body systems, the dependence is also expected to be exponential, but a general formula is yet to be determined.  

Periodic boundary conditions (PBCs) are a specific case of twisted boundary conditions (TBCs) \cite{Byers:1961zz} at the faces of the cubic volume.  These `twist' conditions can be parametrized by a vector of angles $\theta_i$ at each boundary, with range $0\le \theta_i<2\pi$, such that 
\begin{equation}\label{eqn:twist bc}
\psi({\bf x} + {\bf n}L)=e^{i\vec{\theta}\cdot{\bf n}}\psi({\bf x})\ .
\end{equation}
Equation~\eqref{eqn:twist bc} shows that $\theta_i=0$ corresponds to PBCs, while $\theta_i=\pi$ gives anti-periodic boundary conditions (aPBCs).  In LQCD, TBCs are equivalent to introducing a background $U(1)$ gauge field imposed on the quarks, subsequently endowing them with an arbitrary momentum dependent on the twist angle \cite{Guagnelli:2003hw,Bedaque:2004ax}.  With TBCs momentum states are no longer restricted to the discrete modes within a box with PBCs, and therefore calculations with different twist boundary conditions will give rise to different finite-volume corrections.  As initially found within condensed matter calculations, averaging results with different twist angles significantly cancels finite-volume effects \cite{Lin:2001zz}.  This has motivated the use of `twist averaging' in LQCD calculations to reduce the finite-volume dependence in hadronic masses \cite{Lehner:2015bga} and more recently to calculations of phases of nuclear matter in dense astrophysical environments \cite{Schuetrumpf:2015nza}.
  
Because of the non-linear nature of interactions in the non-perturbative regime between quarks and gluons and also between nucleons, twist averaging does not completely eliminate finite-volume effects.  To what extent it does eliminate finite-volume effects is an open question, and most certainly depends on the nature of interactions.  In \cite{Briceno:2013hya} the behavior of finite-volume corrections for the two-body system was investigated for specific sets of twist angles.  It was found that certain linear combinations of twist angles indeed reduce significantly finite-volume effects.  Just as important, it was shown that a particular set of twist angles ($\theta_i=\pi/2$), dubbed 'i-periodic', also significantly reduced the leading-order exponential dependence of the finite-volume.  

In this paper we extend the work done in \cite{Briceno:2013hya} to three-body systems.  Except for particular three-body limits (see, e.g., \cite{Meissner:2014dea}), analytic calculations in this regime are not possible and we utilize the NLEFT formalism to perform our calculations.  In this case, nucleons are the relevant degrees of freedom, not quarks, and therefore twist angles are applied to nucleon state functions directly.  We perform a detailed statistical analysis of our calculations, accounting for and propagating all relevant systematic errors in our extrapolations. From our analysis we find the analogue of `i-periodic' angles for the three-body system, which not only reduce finite-volume effects, but cancels exactly the leading-order FV contribution.  

Our paper is organized as follows: in Section~\ref{sect:twists for N bodies} we discuss the formalism for twisted boundaries on general terms.   We then apply this formalism to the NLEFT algorithms which we coded specifically for the two- and three-body systems.   We describe this in detail in Section~\ref{sect:applying twists}. Included in this same section is an enumeration of sources of systematic errors due to lattice artifacts and the finite volume, and a detailed discussion of our error analysis used to propagate errors. We present results of the  two-body (deuteron) system and the three-body (triton) system in Sections~\ref{sec:intro-two_body} and~\ref{sec:three-body}, respectively.  Finally, we recapitulate our findings and discuss possible future applications in Section~\ref{sect:conclusion}.

\section{Twisted boundary conditions for $N$-body systems\label{sect:twists for N bodies}}
Assuming a non-relativistic $N$-body system, this system's state can be written as a linear combination of tensor products of the individual particle states which include their internal quantum numbers, 
\begin{equation*}
	\ket{\N{1}} \otimes \cdots \otimes \ket{\N{N}}\ ,
\end{equation*}
where $I_i$ and $S_i$ refer to the isospin and spin of the $i^{th}$ nucleon, respectively.  
To realize such states computationally, an appropriate bra basis, either in configuration or momentum space, must be used.  For example, momentum space calculations using a finite cube with PBCs would utilize a discrete momentum basis $\vec{p}_n=2\pi\vec{n}/L$, where $\vec{n}$ represents a triplet of integers.  With an eye towards our lattice simulations presented in later sections, we adopt a discretized coordinate basis: $\vec r \mapsto a \vec n$, where $a$ is the lattice spacing between lattice nodes.  We denote this discretized basis as $\ket{\vec{n}}$.  We stress that the results of this section do not depend on the choice of basis, however.  

\par
A spatial cutoff $L$ is introduced by limiting the basis to a cubic box of volume $L^3$ with application of particular boundary conditions at the faces of the cube. This introduces finite-volume effects (errors) that can only be removed by extrapolating to infinite volume, $L\rightarrow\infty$. In this paper, we focus our analysis mainly on these finite volume effects for the lattice. Objects defined inside the box, such as matrix elements of operators using the discretized basis $\ket{\vec{n}}$, will be denoted with a subscript $L$ to differentiate them from their infinite-volume counterparts. 

The most commonly used boundary conditions are periodic boundary conditions (PBCs), where the wave function is periodically continued outside of the box\footnote{This kind of behavior can also be defined for a finite continuous space.}, which in turn produces images of the wave function outside of the original cubic volume.  Periodic boundary conditions are just a subset of the more general twisted boundary conditions (TBCs) defined as follows:
\begin{align}
	\tensor*[_i]{
		\braket{ \vec{x}_i + L \ \vec n | \Psi_L }}{_N} 
	= 
	\tensor*[_i]{
		\braket{ \vec{x}_i  | \Psi_L }}{_N} e^{i \vec \phi_i \cdot \vec n},
	\qquad
	\forall \ \vec x _i & \in L^3 \ , \ \forall \ \vec n \in \mathbb Z ^3\ .
\end{align}
The variable $ \vec \phi _i \in \mathbb R ^3$ represents the twisted boundary angle of the $i^{th}$ particle with components for each spatial direction. Suppressing the spin and isospin components, one can now define a basis for the discretized box (with $N_L$ nodes in each spatial direction where $L = a N_L$) which satisfies TBCs and is useful for computing matrix elements,   
\begin{align}
	\nonumber
	 \ket{ \mathcal{N}_{{i,L}}}
	 &=
	 \sum_{\vec{n} \in \mathbb{Z}^3}
	 	\mathcal{N}_{i,L}(\vec{n}) \ket{\vec{n}}_i
	=
	\frac{1}{\sqrt{M}^3}
	\sum_{ \underset{ \vec{n} \in N_L^3 }{ \overset{   \vec{m}  \in  \mathbb{Z}^3 }{}}}
	 	\mathcal{N}_{i,L} (\vec{n} +N_ L \vec{m}) \cdot \ket{\vec{n} + N_L \vec{m}}_i
	\\ \nonumber
	&=
	\frac{1}{\sqrt{M}^3}
	\sum_{ \underset{ \vec{n} \in N_L^3 }{ \overset{   \vec{m}  \in  \mathbb{Z}^3 }{}}}
	 	\mathcal{N}_{i,L} (\vec{n})  \ e^{ i \vec{\phi}_{ i } \cdot \vec{m} } \
		 e^{ - i \vec{\phi}_{ i } \cdot \vec{n}/N_L } \  e^{ i \vec{\phi}_{ i } \cdot \vec{n}/N_L } \
		\ket{\vec{n} + L \vec{m}}_i
	\\ \label{eq:derive_twist_basis}
	&=:
	\sum_{\vec{n} \in N_L^3}
	 	\tilde{\mathcal{N}}_{i,L} (\vec{n})
		\ket{\vec{n}}^{\vec{\phi}_i}_i  \ ,
\end{align}
where
we have used the following definitions, 
\begin{align*}
	M^3 & := \sum\limits_{\vec{m} \in \mathbb{Z}^3} 1 \, ,
	&
	\tilde{\mathcal{N}}_{i,L} (\vec{n}) & := \mathcal{N}_{i,L} (\vec{n}) \cdot  e^{- i \vec{\phi}_i / N_L}\, .
\end{align*}
All phases can be absorbed by the newly defined basis states $\ket{\vec{n}}^{\vec{\phi}_i}_i$ and the wave function $\tilde{\mathcal{N}}_{i,L} (\vec{n})$. This new `twisted basis', which we denote collectively by $\{ \ket{\vec{n}}^{\vec{\phi}_i}_i \}$, can be understood as a grid of vectors multiplied by a phase associated with the different images of the original cube:
\begin{align}
	\label{eq: twist basis}
	\ket{\vec{n}^{\vec{\phi}_i}}_i  := \frac{1}{\sqrt{M}^3} \sum_{\vec{m} \in \mathbb{Z}^3}
		e^{ i \vec{\phi}_{ i  } \cdot (\vec{n} + N_L \vec{m})/N_L }  \cdot \ket{\vec{n} + N_L \vec{m}}_i
\end{align}

Note that this description of twisted boundary states is in general different from the twisting convention used in Lattice QCD. In Lattice QCD one directly applies twists to the quarks and, in principle, there are different associated twist angles for each quark flavor.  In our case the twists are directly applied to the configuration space coordinates of the nucleons. Thus the number of twists is directly related to the number of nucleons -- for $N$-nucleons one could choose $N$ different twist angles. 

\begin{figure}[htb]
\begin{subfigure}{.5\textwidth}
  \includegraphics[width = .85\textwidth]{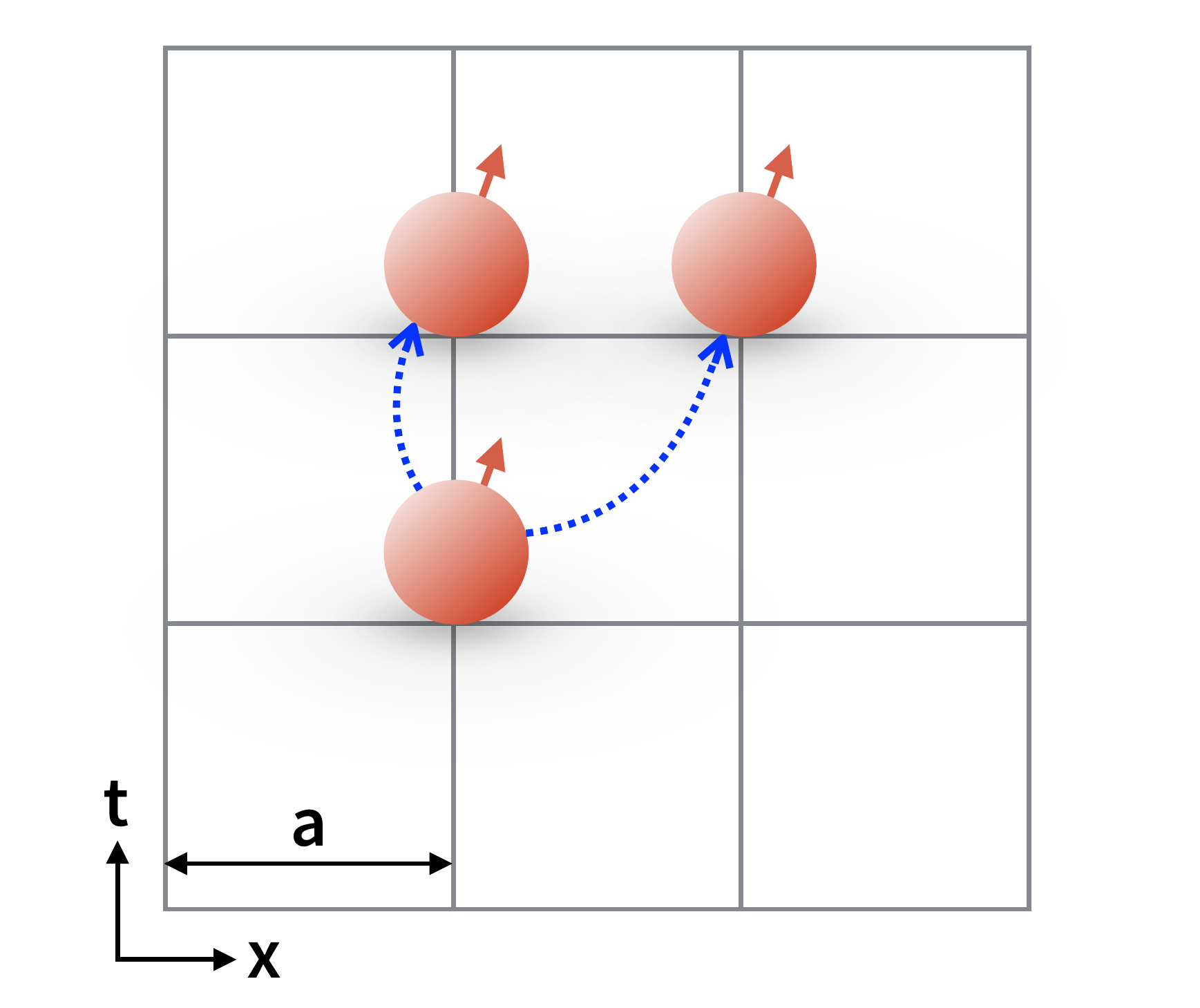}
  \caption{\label{fig:particle_hop_1}Twist at boundaries}
\end{subfigure}%
\begin{subfigure}{.5\textwidth}
  \includegraphics[width = .85\textwidth]{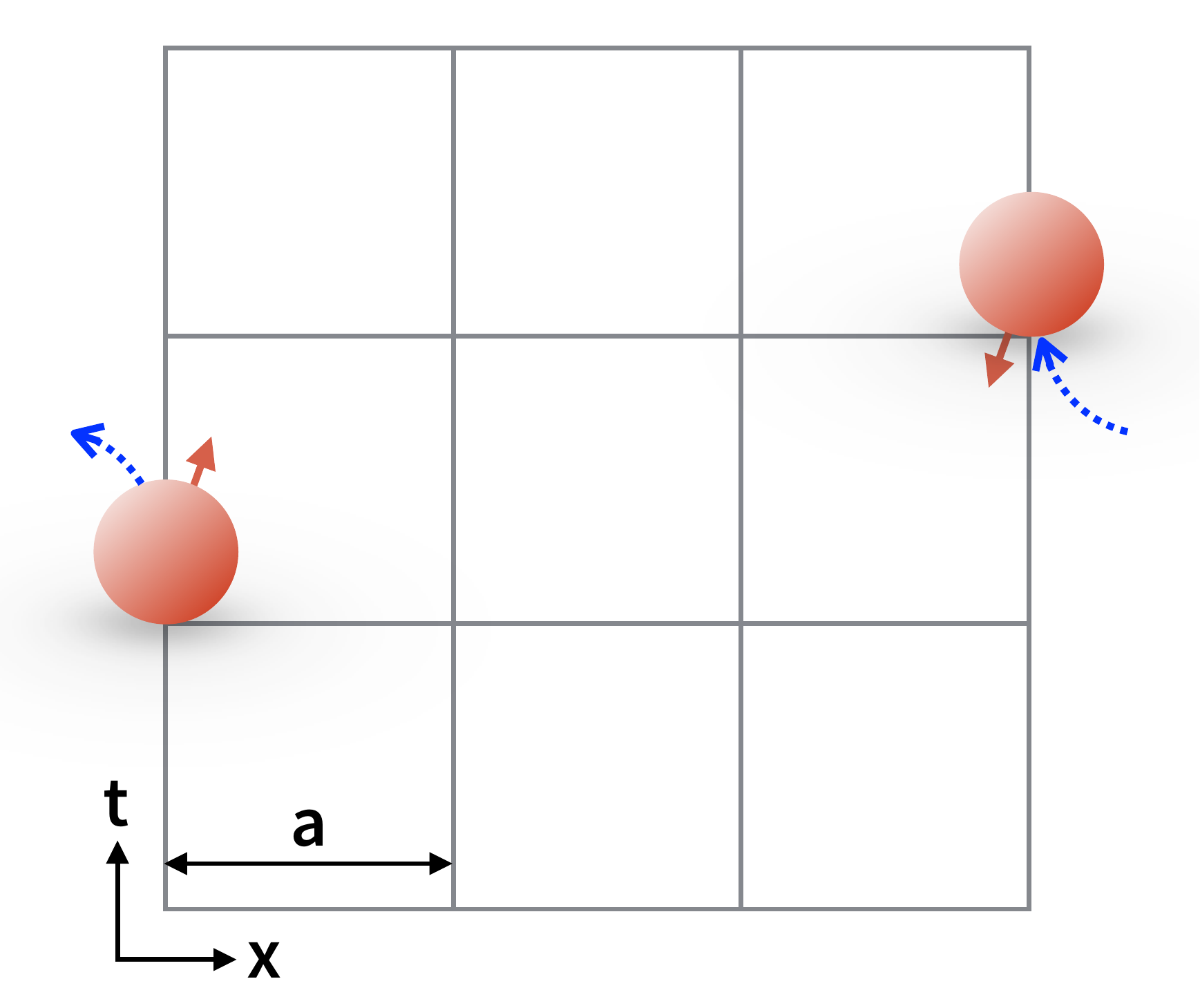}
  \caption{\label{fig:particle_hop_3}Twist at boundaries}
\end{subfigure}%
\\
\begin{subfigure}{.5\textwidth}
  \includegraphics[width = .85\textwidth]{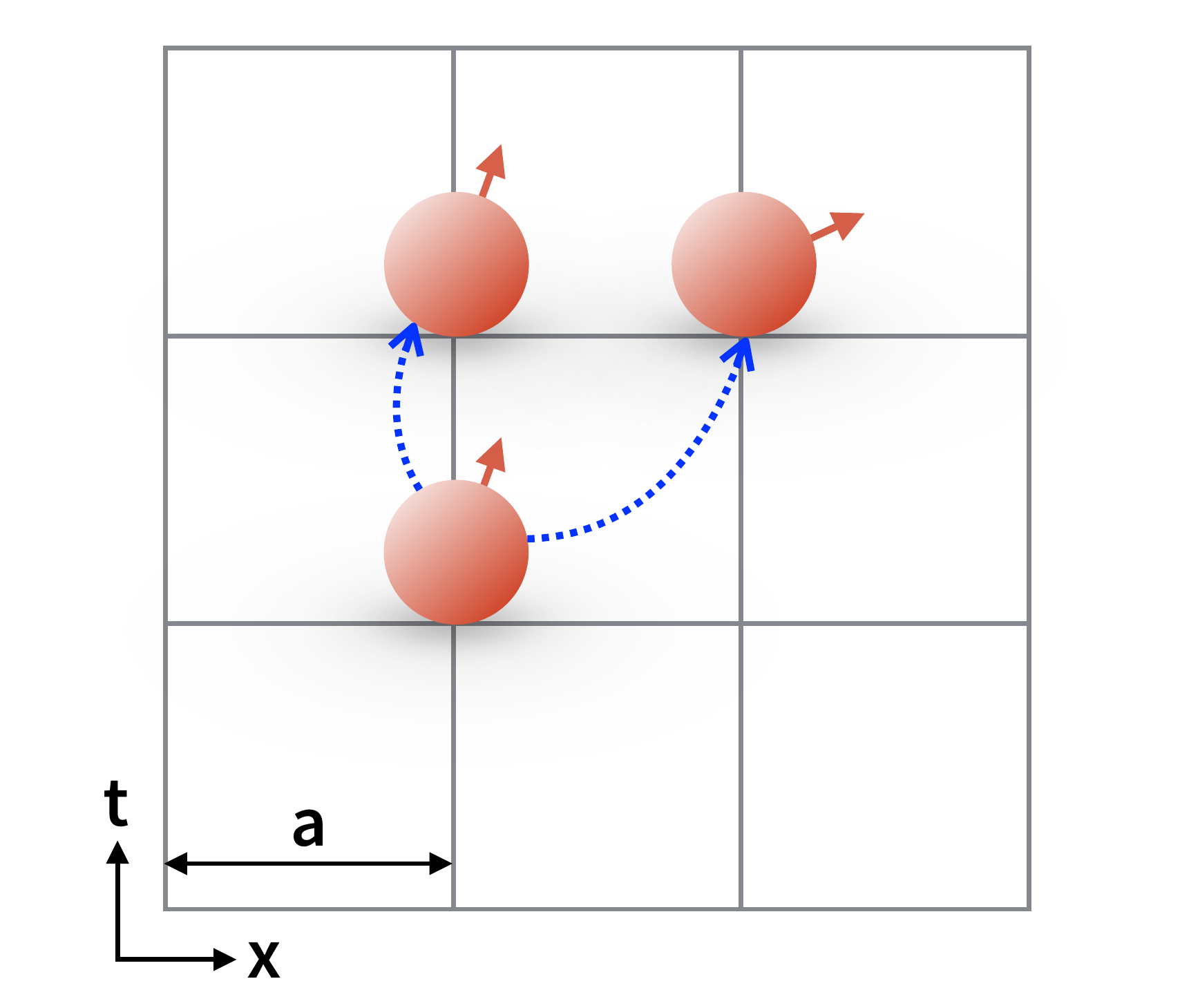}
  \caption{\label{fig:particle_hop_2}Twist at each 'hop'}
\end{subfigure}%
\begin{subfigure}{.5\textwidth}
  \includegraphics[width = .85\textwidth]{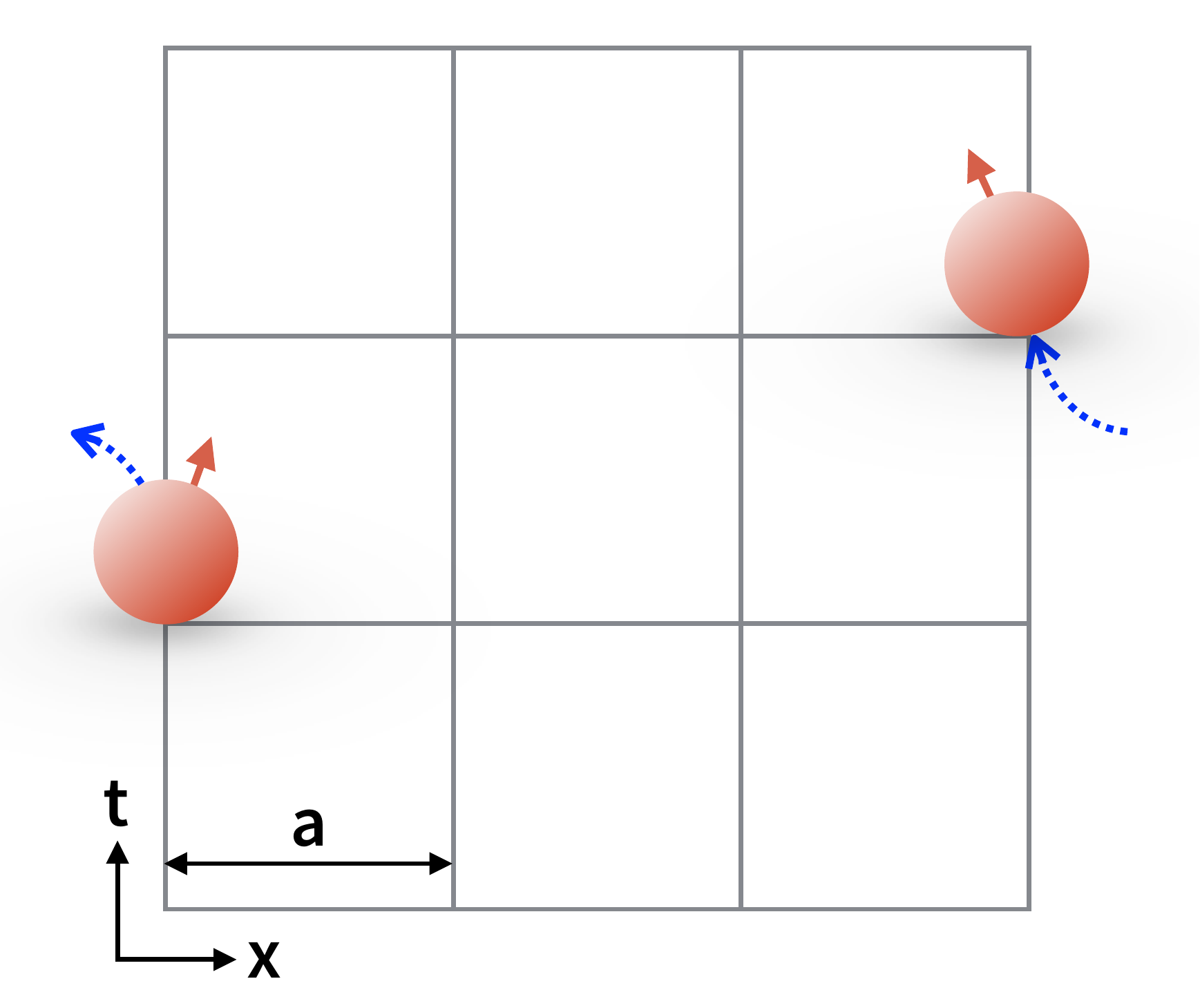}
  \caption{\label{fig:particle_hop_4}Twist at each 'hop'}
\end{subfigure}%
\caption{\label{fig:particle_hop}
A demonstration of the different choices of twisted boundary basis for a simple 2-dimensional lattice with anti-periodic boundary conditions in spatial direction ($\phi = \pi$). The phase of the particles is represented by the direction of the red arrow.}
\end{figure}

We note that for numerical simulations, there is some freedom in how one implements TBCs.  Typically TBCs are applied at the boundaries of the box where the application of the phase $\phi_i$ only occurs when a particle passes the boundary.  We have instead chosen to apply twists
incrementally $\propto \phi_i / N_L$ each time a particle changes its coordinates within the volume.  In this manner, the accumulation of the entire phase $\phi_i$ also occurs when a particle passes a boundary.  Though we have implemented this latter choice, we stress that both choices are equivalent when it comes to calculated observables.  In figure~\ref{fig:particle_hop}  we provide a schematic comparison of our twist basis to the case where twists are applied at the boundaries only.  
In the demonstrated case of anti-periodic boundaries, one has that the wave function flips its sign after a shift in $L$, $ \psi ( \vec x + \vec e L ) = - \psi ( \vec x )$, $| \vec e | = 1$.
The upper two diagrams (\ref{fig:particle_hop_1}, \ref{fig:particle_hop_3}) use a basis where the boundary conditions are just applied at the edges of the spatial box. For this basis, the wave function only changes its phase when the particle `hops' outside of the box \eqref{fig:particle_hop_3} but stays the same if moving within the box \eqref{fig:particle_hop_1}. The lower diagrams (\ref{fig:particle_hop_2}, \ref{fig:particle_hop_4}) use the previously defined twisted boundary basis where a partial phase is applied each time a particle changes its spatial direction.  After $N_L$ steps in the same spatial direction, the accumulated phase becomes $\pi$ and thus the wave function changes its sign. This behavior also holds true at the boundaries, however, it depends on the direction of the `hops'.

The main reason for including the twisted boundary phases at each particle `hop' is that this procedure ensures translational invariance at each point on the lattice, including the boundaries, and is thus more amenable to the NLEFT formalism.  As a consequence of the translational invariance, the momentum modes in the lattice are well defined,
\begin{equation}\label{eqn:momentum}
	\braket{ \vec{p} \ | \vec{x} \, }^{\vec{\phi}_i}_i
	=
	\exp \left( \frac{2 \pi i}{L} \ \vec{x} \cdot \vec{n}_p  \right) \delta^{(3)} \left( L \vec{p} - 2 \pi \vec{n}_p - \vec{\phi}_i \right)\ .
\end{equation}
The delta function in equation~\eqref{eqn:momentum} shows that the allowed momenta inside the box are shifted by the twists for each particle,
\begin{equation}
\vec{p} = \frac{ 2 \pi \vec{n}_p + \vec{\phi}_i}{L}\ .
\end{equation}
It is therefore possible to induce a non-zero center of mass (CM) energy for the zero momentum modes, which is proportional to the twist angles,
\begin{equation}
	E^{(CM)}_{0}
	=
	\frac{ \vec P_{0} ^2 }{2 M_{CM} } 
	= 
	\frac{ 1 }{2 M_{CM} } 
		\left(
			\sum\limits_{i=1}^N \vec \phi _i
		\right)^2\ .
\end{equation}
This CM motion must be accounted for when comparing calculations of the relative binding energies of $N$-body systems.  We note that twist angles subject to the constraint that $\vec \phi _1 + \cdots \vec \phi _N=0$ will induce no CM motion.

To conclude this section, we consider the matrix element of some arbitrary operator $O$ using the basis in equation\eqref{eq: twist basis},
\begin{equation}\label{eqn:M1}
	\braket{ \vec m _1^{\vec \phi _1} , \cdots ; \vec m _N^{\vec \phi _N} | O| \vec n _1^{\vec \phi _1} , \cdots ; \vec n _N^{\vec \phi _N} }\ .
\end{equation}
As the operator $O$ can be written in terms of products over creation and annihilation operators, it suffices to consider the following term,
\begin{align}\label{eqn:hopper}
	 \nonumber
	\sum_{\vec{n}^\prime \in \mathbb{Z}^3}
	a_i^\dagger( \vec{n}^\prime + \vec{l} \ ) a_i (\vec{n}^\prime ) \ket{ \vec{n}^{ \phi_i} }_i
	&= 
	\frac{1}{\sqrt{M}^3} \sum_{\vec{m} \in \mathbb{Z}^3}
		e^{ i \vec{\phi}_{ i  } \cdot (\vec{n} + N_L \vec{m})/N_L } 
		\ket{ \vec{n} + \vec{l} + N_L \vec{m} }_i
	\\
	&=
	 \ket{ (\vec{n} + \vec{l})^{ \phi_i}}_i
		e^{ - i \vec{\phi}_{ i  } \cdot   \vec{l}  /N_L }\ .
\end{align}
This term is off-diagonal in the basis of creation and annihilation operators, and represents a `hopping term' from site $\vec{n}$ to site $\vec{n}+\vec{l}$.  Equation~\eqref{eqn:hopper} explicitly shows how a particle picks up an incremental phase through such a translation between sites.   More generally, any operator $O$ with non zero off-diagonal matrix elements in creation and annihilation operators will be modified by a phase within the `twisted basis',
\begin{multline}
	\label{eq: operator twists}
	\braket{
		\vec m _1^{\vec \phi _1} , \cdots ; \vec m _N^{\vec \phi _N} | O | \vec n _1^{\vec \phi _1} , \cdots ; \vec n _N^{\vec \phi _N} 
	}
	\\
	=
	\braket{
		\vec m _1^{\vec 0} , \cdots ; \vec m _N^{\vec 0} | O | \vec n _1^{\vec 0} , \cdots ; \vec n _N^{\vec 0} 
	}
	\exp
	\left(
		i \sum \limits _{i = 1}^N \vec \phi_i \cdot ( \vec n _i - \vec m _i ) 
	\right)\ .
\end{multline}
Therefore the off-diagonal $N$-body matrix element with TBCs are equal to the $N$-body matrix element with PBCs multiplied by a phase that depends on the twist angles.
It is important to stress that these matrix elements still represent a hermitian matrix if the evaluated operator is hermitian as well. In other words, twisted boundaries do not induce (extra) sign oscillations.

\section{Applying twists within the NLEFT formalism\label{sect:applying twists}}
To study the effects of the finite volume on two- and three-body systems, we perform calculations on a discretized space-time lattice within a cubic volume.  Our implementation follows closely that of the Nuclear Lattice Effective Field Theory (NLEFT) formalism.   Though the NLEFT algorithm is well documented (for a review of NLEFT, see~\cite{Lee:2008fa}), we provide a cursory description of our algorithm mainly to point out differences with past NLEFT calculations and to describe our implementation of TBCs within the NLEFT formalism.   

\subsection{Twists on the transfer matrix $\mathcal M$}
To obtain results of nuclear observables on a lattice, one computes the trace of products of the transfer matrix, $\mathcal M$, which in our case is identified with the chiral interaction of nucleons.  
Formally, the transfer matrix in Euclidean time is given by the normal-ordered exponential of the corresponding effective Hamiltonian,
\begin{equation}
	\mathcal M := \ : \exp \left( - H a_t \right) \ :\ .
\end{equation}

The spectrum of H can be ascertained from eigenvalues of the transfer matrix $\mathcal M$,
\begin{equation}
\mathcal M  \ket{\psi_n} = \epsilon_n \ket{\psi _n}\quad,\quad\epsilon_0 > \epsilon_i \ , \ \forall \ i > 0\ .
\end{equation}
In particular, the ground state energy $E_0$ of the system is related to the largest eigenvalue of $\mathcal M$, which we denote as $\epsilon_0$, and can be obtained through the following logarithmic derivative  
\begin{equation}
	E_ 0 = - \frac{\log(\epsilon_0)}{a_t}\ .
\end{equation} 

The Lagrangian which generates the transfer matrix contains the leading order chiral contact interactions given in, for example, \cite{RevModPhys.81.1773}. Effectively one obtains a two-body force as well as the one-pion exchange at leading order,
\begin{equation}
	V_\chi^{(LO)} (q)
	=
	V_{NN}^{(LO)} (q)
	+ 
	V_{\pi N}^{(LO)}(q)\ ,
\end{equation}
where
\begin{equation}
	V_{\pi N}^{(LO)}(q)
	=
	-
	\left(\frac{g_A}{2 f_\pi}\right)^2 \frac{ \left( \vec \sigma _1 \cdot \vec q \right)   \left( \vec \sigma _2 \cdot \vec q \right) }{q^2 + m_\pi^2}
	\ .
\end{equation}
Here the nucleon mass and the pion mass are set their physical value $m_N = 938.92 $ MeV and $m_\pi = 134.98 $ MeV. The pion decay constant is $f_\pi=92.2 $ MeV and the axial coupling has a strength of $g_A = 1.29$ respecting the Goldberger-Treiman discrepancy for representing the strong $\pi N N$-coupling. Furthermore the momentum $\vec q=\vec p ^\prime - \vec p$ is the nucleon momentum transfer.
In this work, the contact potential was implemented using a gaussian-like smearing in momentum space similar to the one used in \cite{Borasoy:2006qn}, 
\begin{equation}
	V^{(LO)}_{NN} ( q ) 
	= 
	\left(
	c_{SU4} + c_{I} \ \tau_1 \cdot \tau _2 + c_S \ \vec \sigma _1 \cdot \vec \sigma _2 + c_{SI} \ \tau_1 \cdot \tau _2 \ \vec \sigma _1 \cdot \vec \sigma _2
	\right)
	e^{ - b_4 q^4 }\ .
\end{equation}
Furthermore, the coefficients $ ( c_{SU_4}, c_{S}, c_{I}, c_{SI})$ were related to each other through the leading order singlet and triplet coefficients $ C_{S}$ and $C_{T} $ when evaluating nucleonic matrix elements,
\begin{align}
	c_{SU4} &= \frac{1}{16} \left( 3 c_{S} +  3 c_{T} \right)
	&
	c_S &= \frac{1}{16} \left( -3 c_{S} +  c_{T} \right)
	\\ \nonumber
	c_I &= \frac{1}{16} \left( 3 c_{S} -  c_{T} \right)
	&
	c_{SI} &= \frac{1}{16} \left( - c_{S} -  c_{T} \right)\ .
\end{align}
\begin{table*}[tb]
\caption{\label{tab:computation parameters}Numerical values of parameters used in our simulations.}
\begin{ruledtabular}
\begin{tabular}{c c c c c c}
	$1/a_L$	& $1/a_T$ & $ c_{S}$ & $c_{T} $ & $\Delta ^{(n)} $ & $b_4$ 
	\\
	MeV	& MeV	& $  10^{-5} $ MeV$^{-2}$	& $  10^{-5} $ MeV$^{-2} $ &  $\mathcal O (a^{2 \cdot n})$ & MeV$^4$
	\\ \hline
	$100 $	& $ 150 $	& $ -4.2000 $	& $ -6.0513 $&	$ \mathcal O (a^{2 \cdot 4}) $ & $0.07$ 
\end{tabular}
\end{ruledtabular}
\end{table*}
The contact interactions were  fitted to reproduce the deuteron binding energy as well as the $^3 S_1$ scattering length. We tabulate their values, as well as other parameters relevant to our simulations, in table~\ref{tab:computation parameters}.  To reduce the dimensionality of the problem, the spin breaking part of the pion exchange was assumed to be small and computations with this part were done for one specified spin channel only. This induced a small error when comparing to the `experimental result' at the order of $0.05 $ MeV for the deuteron. 

Because of our `low-order' interaction, we do not expect to have perfect agreement for the three-body energy levels when compared to experiment. However, since the goal of this paper is to emphasize the dependence of the binding energy of few-body systems on FV corrections, this level of simplicity for the nucleon interactions is sufficient.   As such,  one should compare calculated energy levels in a fixed volume to their converged results for large (infinite) volumes instead of to the experimental results themselves.

Furthermore, at this order our potential does not contain any derivatives acting on the nucleon coordinates and therefore does not induce translations on the nucleon states.   The inclusion of TBCs is therefore realized by implementing equation~\eqref{eq: operator twists} for the kinetic hamiltonian operators only. 

As a final comment, we point out that the normal ordering of the transfer matrix for two nucleons $\mathcal{M}^{(2)}$ is exact at order $a_t^2$,
\begin{equation}
	\mathcal M^{(2)} = \mathbbm 1 - a_t  \left( H_0^{(1)} + H_0^{(2)} + V^{(1,2)}\right)  + a_t^2 H_0^{(1)} H_0^{(2)} \ .
\end{equation}
To identify the CM motion of such a system, one can rewrite the absolute momenta of the individual particles as combinations of the CM momentum $\vec{P}$ and the relative momentum $\vec{q}$,
\begin{multline}\label{eqn:Pq}
	\mathcal M^{(2)}  = 
	\mathbbm{1} - a_t \left( H_0^{(rel)} + H_0^{(CM)} + V^{(rel)} \right)
	  +  a_t^2 \left( \frac{1}{4} \left(H_0^{(rel)} + H_0^{(CM)} \right)^2 - \left(\frac{\vec{P} \cdot \vec{q}}{M_{CM}} \right)^2\right)\ .
\end{multline}
Equation~\eqref{eqn:Pq} shows that the term of second order in $a_t$ couples CM motion to relative motion for non-zero CM momenta. Thus the procedure of subtracting the CM motion from the computed spectrum is more complicated. If one computes the spectrum of a two-nucleon system using general twisted boundaries, the energy eigenvalues of the transfer matrix are shifted by the non-zero CM contributions generated by these twists. This also holds true for more general $N$-body systems as well.  For this current work we avoid the extra complication of non-zero CM coupling by utilizing twists that induce zero CM motion, which is determined by the following constraint,
\begin{equation}
	\sum \limits _{n=1}^N \vec \phi _n = 0\ .
\end{equation}

\subsection{\label{sec:error_analysis}Identification of Systematic Errors and Description of Error Analysis}
Nuclear lattice EFT calculations employ Monte Carlo methods to estimate the ground state energy of $N$-body systems.  Because such methods are intrinsically stochastic, the extracted energies have an associated statistical uncertainty.  In our case, because the dimensions of our systems are so small, we can extract our energies via direct diagonalization of the transfer matrix.  Our energies therefore have no statistical uncertainty. 

Nevertheless, our results are not completely free of `errors', as there still exists sources of theoretical and systematic uncertainties which can induce an effect on the final result. We enumerate these sources here and discuss each in turn below:
\begin{itemize}
	\item Finite-volume effects
	\item Discretization errors
	\item Numerical/rounding errors
	\item Uncertainties associated with the fitting of lattice parameters (LECs on the lattice for NLEFT)
\end{itemize}
Since the aim of our study is the analysis of the FV dependence of the ground state binding energy, the uncertainties associated with the fitting procedure of LECs are neglected. This also holds true for discretization errors which are connected to the implementation of the derivatives and the fitting of the LECs. In LQCD, when one wants to rigorously compute physical observables, one also has to take the $ a \rightarrow 0$ continuum limit.  Here a careful accounting of the discretization errors is needed to perform a robust extrapolation.  In NLEFT this procedure is more complicated since the interactions themselves are cutoff dependent for a given order in the effective expansion. In our case we do not perform a continuum analysis since, again, we only focus on comparing the computed energies in a finite volume to their infinite volume counterparts.  In all calculations we use the same lattice spacing.

Our numerical errors are associated with the solving procedure only, which involves a Lanczos-like iteration for diagonalizing the transfer matrix and obtaining eigenvalues. This method of solving does not introduce any statistical errors. These numerical errors are of the order $\epsilon=e^{-Ea_t}\le10^{-5}$, which corresponds to an energy error budget of 
\begin{align}
	\delta E _\epsilon
	& \leq
	\frac{ \delta \epsilon}{ a_T \epsilon }
	\simeq
	0.002 \ \text{MeV}\ .
\end{align}

In contrast to the previous errors, which are volume independent,  we note that the FV energy corrections are only asymptotically diminishing if the potential vanishes within the cubic volume. Formally the FV needs to be of size $ L/2 \gtrsim R$, where $V(R) \simeq 0$. Furthermore, for small boxes, next to leading order (NLO) FV corrections become more relevant. 
As an example, the functional form of the leading-order FV expression of the energy shift for two-body states,
\begin{align}
	E_L - E_\infty 
	&=
	\sum \limits _{n = 0}^{\infty}
		\Delta E^{(n)}_L \ ,
	&
	\Delta E^{(LO)}_L
	:=
	\Delta E^{(0)}_L 
	\propto 
	\frac{e^{- \kappa L}}{L} \ ,
\end{align}
where $\kappa^2 = - m_N E_\infty > 0$ is the two-body binding momentum, does not perfectly describe our numerical results, particularly at small volumes, since the complete energy shift includes higher order corrections described by several exponential functions of different exponents and amplitudes \cite{Luscher:1986pf}. Since one of our objectives of this study is to use calculations within small volumes to extract infinite volume observables, we must explicitly take into account the errors from neglecting NLO (and higher) FV effects.  We do this by estimating the size of the NLO FV systematic error, $\Delta E_L^{(NLO)}(L)$, and inflating our binding energy uncertainties by this amount when performing our fitting and error analysis.
\begin{equation}
	\Delta E_L^{(NLO)}(L,\phi)
	\mapsto
	\delta ( \Delta E_L^{(LO)}(L, \phi) )
	\ .
\end{equation}
We stress that $\Delta E _L$ denotes the analytic form of the finite volume corrections, while quantities labeled with a small delta, $\delta$, are treated as uncertainties of the computation and fitting procedure.

In principle, this NLO FV term should be interpreted as a weight for the fitting procedure which increases the relevance of data points at larger box sizes (where NLO FV effects become less important). 

With the sources of errors described above, we parametrize the total \emph{uncertainty} for the binding energy, $\delta(E_L - E_\infty)$, by the following terms,
\begin{equation}
	\delta( E_L - E_\infty) \simeq 
	\delta( \Delta E_L^{(LO)}(L,\phi) ) + \delta E_\epsilon
	=:
	\delta E_L(L, \phi) + \delta E_\epsilon 
\end{equation}
Furthermore, since the general twist dependence of the NLO FV corrections is not known, it will be assumed that these errors can be estimated by the periodic boundary result,
\begin{equation}
	\delta E_L(L, \phi) + \delta E_\epsilon 
	\simeq 
	\delta E_L(L, \phi) \big | _{\phi = 0} + \delta E_\epsilon 
	=: 
	\delta E (L)
	\ .
\end{equation}
Note that one can assume that the errors associated with this effect might be correlated, e.g., that each data point for a given twist is shifted in the same direction by the NLO FV corrections. 

Our final objective is to extract the infinite volume binding energy $E_\infty$ as well as the coefficients obtained by fitting the leading order FV behavior $\Delta E^{(LO)}_L(L, \phi)$. To estimate the uncertainties of the fitted parameters, we employed a bootstrap-like procedure in our fitting process.  We first performed calculations of binding energies at different values of $L$ and twist angles $\phi$. We designate the collection of such results as $D_0$.  
From $D_0$ we generated $N_{s}$ new distributions $D_i$ by sampling data points within $D_0$ assuming the data points were randomly distributed\footnote{Quantitatively similar results have been obtained for a gaussian distribution and a uniform distribution. The propagated errors of the uniform distribution have slightly more spread.} in the interval of the original error bars $\delta  E(L)$. All of these new distributions were fitted and the newly obtained fitting parameters were used to compute their variation. Thus, the error as well as the mean value of the fitting parameters $F$ were obtained by sampling the new distributions $P(F)$ of fitted parameters (which contain $N_s$ data points),
\begin{align}
	\mu _F 
	&= 
	\int \text{d}F \ F P(F)  \ ,
	&
	\Delta F^{(\pm \alpha)} 
	&\leftrightarrow 
	\hspace{-15pt} \int \limits _{ \mu _F }^{ \mu _F \pm  \Delta F^{(\pm \alpha}) } \hspace{-15pt} \text{d}F \ P(F)  =  \pm \alpha\ .
\end{align}
In our analysis, $\alpha=0.341$ was chosen to give 1-$\sigma$ confidence intervals.
The overall $\chi^2$ per degrees of freedom $\chi_{avg}^2$ is given by an average over all individual fit $\chi^2$ for each fit of distributions $D_i$.
\begin{equation}
	\chi_{avg}^2 := \frac{1}{N_{s}}  \sum\limits_{i=1}^{N_{s}} \chi^2_{d.o.f.}(D_i)
\end{equation}

\section{\label{sec:intro-two_body}Two-Body System: The Deuteron}
As shown in \cite{Luscher:1986pf,Bour:2011ef, Konig:2011ti, Konig:2011nz, Koenig:2013di}, it is possible to analytically compute the FV corrections of the binding energy for a two-body system. 
In general this correction depends on the associated boundary angles $\vec \theta$ in relative coordinates, the box size $L$ as well as infinite volume quantities,
\begin{equation}
	\label{eq:two-body_mass_corrections}
	E_L - E_\infty
	\simeq
	\Delta E_L^{(LO)} ( L, \vec \theta )
	=
	- \mathcal A^{(LO)}
		\frac{e^{ - \kappa L}}{ \kappa L} 
		\sum\limits_{i=1}^3 \cos \left( \theta_i \right)\ .
\end{equation}
Here $\kappa$ is the binding momentum $\kappa^2 = - m_N  E_\infty  > 0$ and $\mathcal A^{(LO)}$ is a numerical amplitude which in general depends on the binding energy and the nucleon mass as well as the angular momentum quantum numbers. Example values for $\mathcal A^{(LO)}$ can be found in \cite{Konig:2011nz}. The boundary angle $\vec \theta$ is defined for the relative system and can be associated with the shift of the relative momentum. Thus the individual nucleon twists $\vec \phi _i$ can be related to $\vec \theta$ by 
\begin{equation*}
	\vec \theta = \frac{ \vec \phi_ 2  - \vec \phi _1  }{2} \ .
\end{equation*}
The exponential dependencies on $L$ are caused by the overlap of the images wave functions with the original wave function \cite{Luscher:1986pf} . The $(\kappa L)^{-1}$ suppressions are related to angular momentum behavior of the bound state wave function for higher partial waves. As can be seen for the two-body case, there exist a certain choice of boundary conditions for which the leading order FV corrections directly vanish -- the `i-periodic' boundaries (iPBCs) defined by $\theta _i = \pi /2 \ \ \forall \ i \in \{1,2,3\}$. The next to leading order corrections in $1/(\kappa L)^2$ and $e^{- \sqrt{2} \kappa L}$ are analytically known,
\begin{equation}
	\Delta E _L^{(NLO)}(L) 
	= 
	\frac{e^{-\kappa L}}{\kappa L} \ 
		\left( \mathcal{A} ^{(NLO)} _1 \ e^{- (\sqrt{2} - 1)\kappa L } + \mathcal{A} ^{(NLO)} _2 \ \frac{1}{\kappa L} 
		\right)
\end{equation}
For error estimation, the error amplitude $\mathcal{A} ^{(NLO)}_i$ is assumed to be of the order of the to be fitted amplitudes $\mathcal{A} ^{(LO)}$, and we therefore set $\mathcal{A} ^{(NLO)}_i {=}\mathcal{A} ^{(LO)}$.
\par
\begin{figure}[htb]
\includegraphics[width = .95\textwidth]{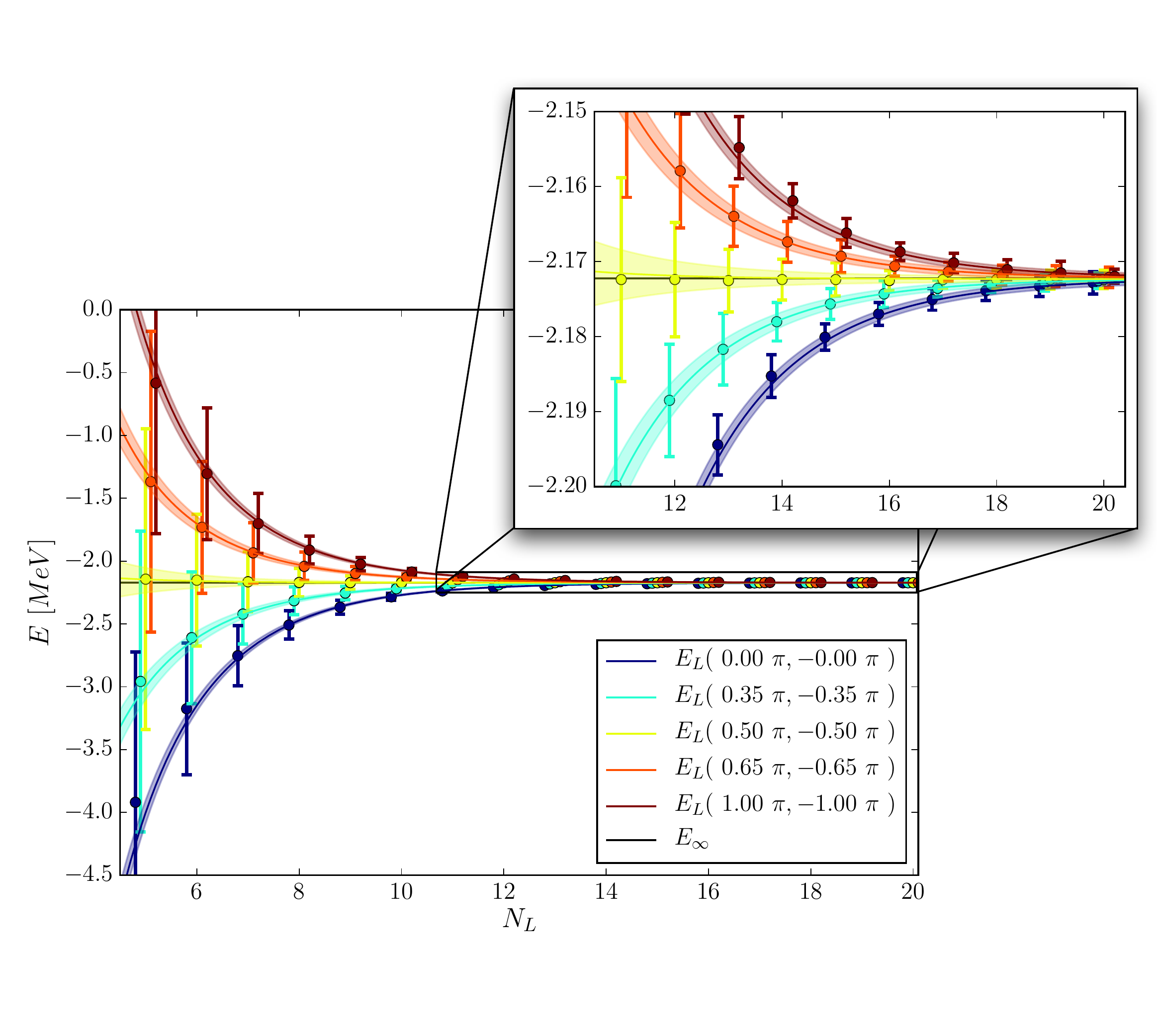}
\caption{\label{fig:deuteron fit}Selected individual fits of two-body binding energy depending on finite volume for $L = a N_L$ with $N_L$ from $4$ to $20$ and $a = 1.97 $ fm. %
 $E_B = \left(-2.172_{-0.001}^{+0.000} \right)$ MeV and $\chi_{avg}^2 = 0.36$ have been similar for each twist configuration according to error propagation. The error bars and error bands correspond to 1-$\sigma$. Data points and bands are slightly shifted in $N_L$ direction for visualization purposes.}
\end{figure}
\begin{figure}[htb]
\centering
\includegraphics[width = .95\textwidth ]{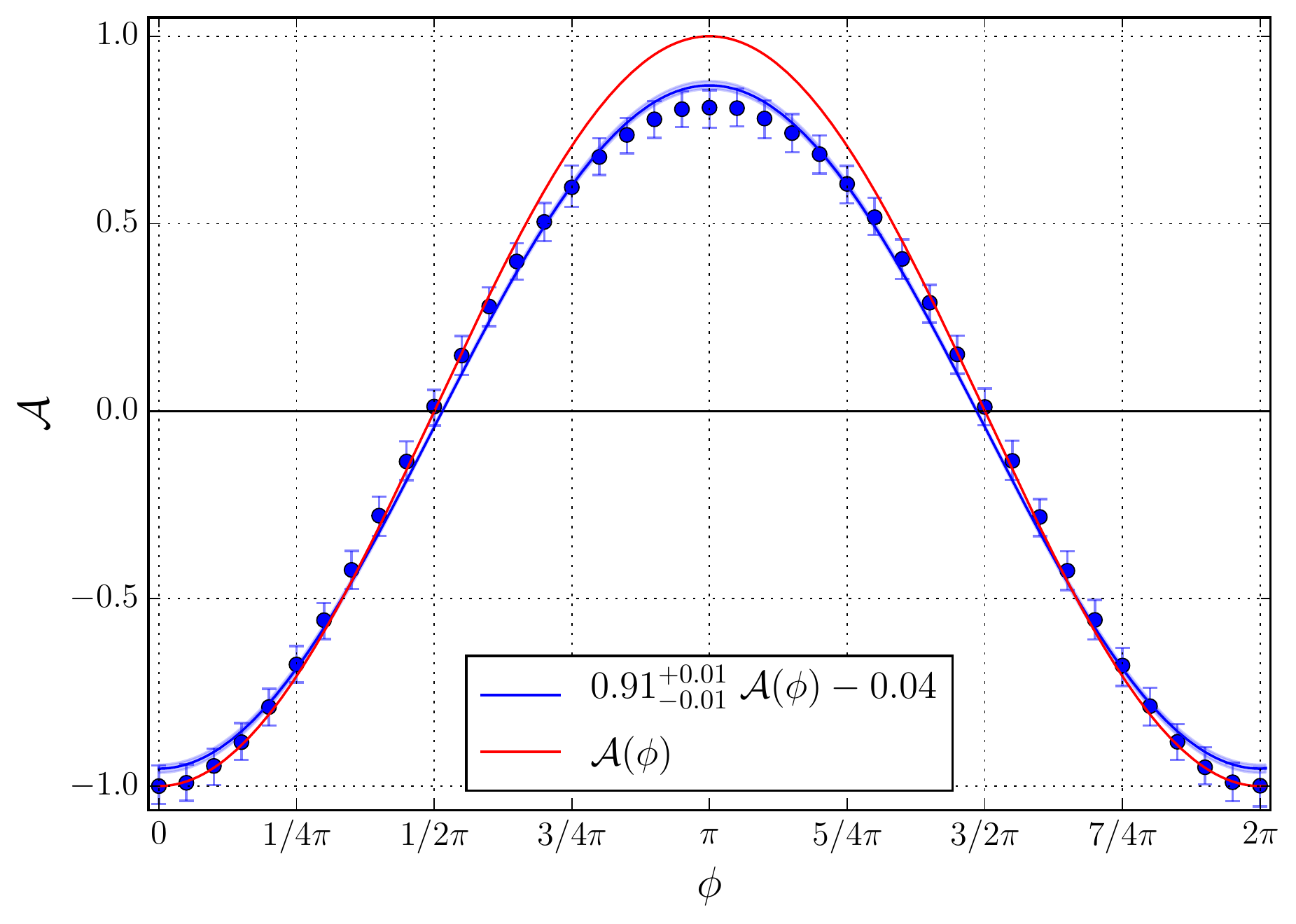}
\caption{\label{fig:coefficient_fit}Fit of two-body coefficient $\mathcal A(\phi)$ depending on relative twist angle $(\phi_1, \phi_2) = (\phi, - \phi)$ with $\chi_{avg}^2 = 0.92$. 
The blue line correspond to a fit of the form $f(\phi) = A \ \mathcal A (\phi) + B$, while the red line is the theoretical prediction $\mathcal A (\phi) = - \cos (\phi)$.
The error bars and error bands correspond to 1-$\sigma$.}
\end{figure}

In this work, the twists have been chosen to be anti-parallel to ensure zero CMS motion. Furthermore each spatial direction is boosted equally by $ \phi _2 = \phi = -\phi _1$ resulting in $\vec \theta = \vec \phi$. Therefore the finite volume energy correction amplitude is proportional to a single cosine factor depending on the twist angle $\phi$,
\begin{equation}
	\Delta E_L^{(LO)} (L,\phi ) = - 3 \, \mathcal A^{(LO)}  \, \frac{e^{-\kappa L}}{\kappa L} \cos ( \phi ) =:  A(\phi) \frac{e^{-\kappa L}}{\kappa L}\ .
\end{equation}
We have performed calculations using 41 different twist boundary conditions, each at multiple volumes $L = a N_L$ with $N_L$ from $3$ to $20$ and a spatial lattice spacing $a = 1.97 $ fm. In figure \ref{fig:deuteron fit} we show a small subset of our twist calculations with their corresponding fits.  The infinite volume binding energy of the deuteron $E_\infty$ as well as the coefficients in front of the exponential $A( \phi)$, shown in figure~\ref{fig:coefficient_fit}, have been extracted from the computed data points using both a constrained fitting procedure where we enforce the same infinite volume $E_\infty$ but different amplitude coefficient for all distributions, and  from individual fitting procedures where we make no constraint on $E_\infty$. The cumulative average of $N_s=1000$ distributions within the data errors results in $\chi^2_{avg} = 0.36$.
The normalized amplitudes $\mathcal A (\phi)  := A(\phi) / A_{max} = - \cos (\phi)$ with $ A_{max} = \max( | A( \phi ) | )$ have been fitted to $ f(\phi) = A \, \mathcal A (\phi)   + B$ (figure \ref{fig:coefficient_fit}).

\begin{figure}[htb]
\centering
\includegraphics[width = .75\textwidth, page=1]{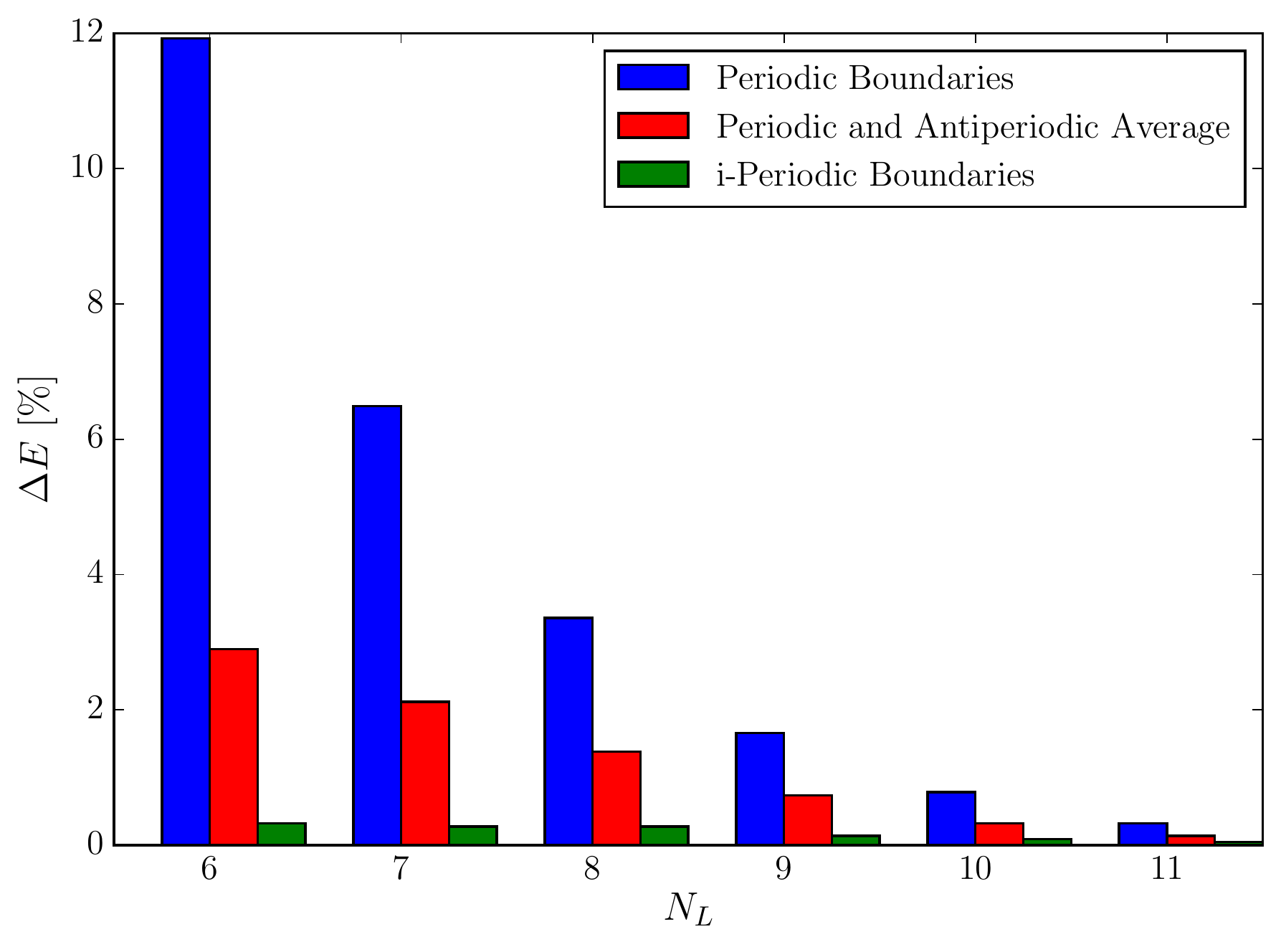}
\caption{\label{fig:convergence_err} Absolute value of relative leading order FV corrections over fit range for different twist averages. This analysis does not take NLO FV corrections into account.}
\end{figure}

\begin{figure}[htb]
\centering
\includegraphics[width = .95\textwidth]{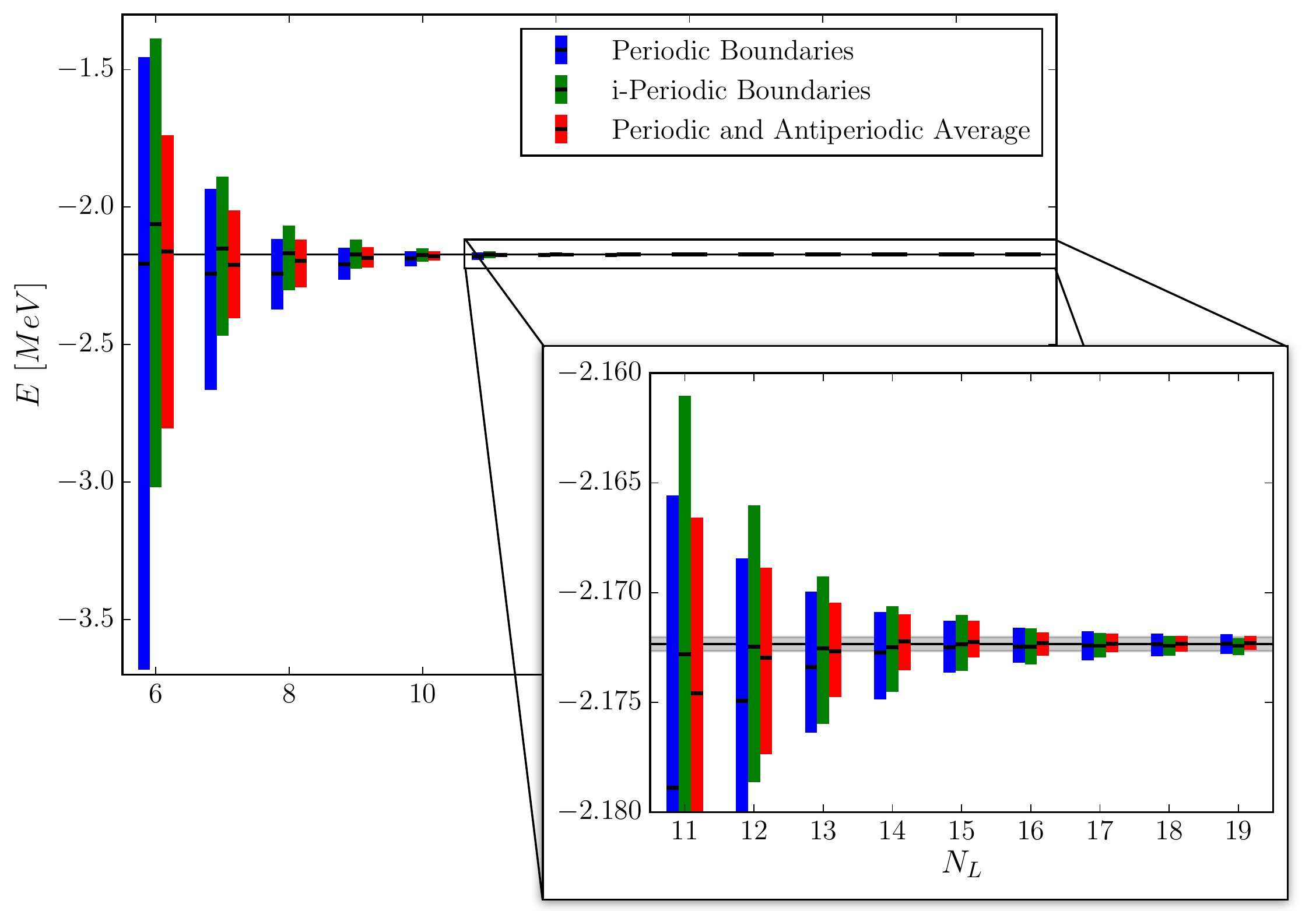}
\caption{\label{fig:deuteron-fit-range}Fit range dependence of deuteron binding energy for different twist averages using NLO FV correction information. The black line corresponds to the mean of the sampled distributions. The error bars correspond to 1-$\sigma$.}
\end{figure}

To emphasize the convergence of the twist averaging, we show in figures~\ref{fig:convergence_err} and \ref{fig:deuteron-fit-range} the extracted binding energies using various fit ranges (from $N_{L,start}=4$ to $N_{L,end} = N_L$) for periodic boundary fits, periodic and anti-periodic constrained fits,  and fits with `i-periodic' twists. In figure \ref{fig:convergence_err}, one can see results obtained without making use of the fitting error propagation -- results which would have been obtained without making explicit use of the NLO FV corrections. 

Furthermore, it is essential to notice that twist averaging as well as iPBCs are quite \emph{insensitive} to the exact form of fitting function. This can also be seen in fig.~\ref{fig:deuteron-fit-range} which shows a similar analysis but includes next to leading order FV corrections.  Note that here the size of the uncertainties for the PBC and aPBC average as well as the uncertainties for iPBCs are smaller than the uncertainties for PBCs only.

As can be seen from figure \ref{fig:convergence_err}, iPBC and aPBC + PBC average results greatly improve the precision of finite-volume results compared to PBC results.  These findings are in complete agreement with those of \cite{Briceno:2013hya} and gives us confidence that our implementation of twists is correctly done.  Since the computational costs grows exponentially with the size of the box, studies computing ground state binding energies can greatly profit using such twists, particularly if similar conditions hold for more complicated $N$-body systems.  

Examination of figure \ref{fig:coefficient_fit}  shows that, for the two-body case with zero induced CM twists, the iPBCs are superior to an average over several twists. This is explained by the fact that the finite-volume corrections are slightly more shifted towards the PBC result, which in turn provides an offset to the extracted energy if twist results are averaged.  This offset can be seen by comparing the data points to the red line in figure~\ref{fig:coefficient_fit}.  Indeed, an integration over all twists points in figure~\ref{fig:coefficient_fit} results in an offset at the order $ \delta E_\infty =  - 0.04 \times \delta \Delta E_L^{(LO)} (L,0)$.  This offset leads to a larger uncertainty in our extracted energies, compared with iPBCs, even when averaging over small sets of twists, as shown in figure \ref{fig:convergence_err}.  A possible explanation could be related to the twist angle dependence of NLO FV effects. Also, the theoretical prediction for the twist dependence is obtained in continuous space. It might be possible that the form of discretization, which does affect momenta on the lattice, most certainly affects twists as well. Further studies are needed to confirm this.

Although our individual fits of $E_\infty$ at different volumes are consistent within uncertainties, as shown in figure~\ref{fig:deuteron-fit-range}, our $\chi_{avg}^2$ are typically below one.  This indicates that our results are correlated and/or our errors have been overestimated.  Indeed, one source of overestimation comes from the fact that we have conservatively assigned a constant error for NLO FV effects, even though these effects themselves depend on twist angles and can be much smaller than our assigned error. 

As a final comment,  we note that with the error budgets that we assign to our sources of errors, we find that the bounds on the uncertainties do not depend drastically on the chosen values of the twists. However the mean value of the infinite volume energy is more accurate for twist averaging and iPBCs compared to PBCs alone.

\section{Three-Body Case: The Triton\label{sec:three-body}}
The exact form of finite-volume corrects for the general three-body case has not been determined to date.  In \cite{Meissner:2014dea}, however, the three-body leading-order FV corrections for three identical particles with PBCs ($\vec \phi _i = 0$ for $i=1,2,3$) in the unitary limit was derived, 
\begin{equation}\label{eqn:meissner equation}
	\Delta E_{L}^{(LO)} ( L , \{ \vec \phi _i = \vec 0 \} )
	=
	\mathcal A ^{(LO)}_{PB} \frac{ \exp \left( - \frac{2}{\sqrt{3}} \kappa L \right)}{ (\kappa L)^{3/2} } 
	\ .
\end{equation}
Though this form is not rigorously applicable for our system (our system is not at the unitary limit), it is sufficient for our analysis we describe below.  We stress that our main conclusions of this section do not depend on the specific FV functional dependence shown in equation~\eqref{eqn:meissner equation}.    

As was done in the two-body case, we assume that the NLO FV corrections come from additional powers of $(\kappa L)^{-1}$ for different partial-wave channels as well as suppressed terms coming from the overlap of diagonally shifted images of the wave functions\footnote{The argument of the wave function gets shifted by one box size in two different directions: \\ $  \psi ( \vec n ) \mapsto \psi ( \vec n + (\vec e _1 + \vec e _2 ) L )$} with the original wave function,
\begin{equation}
	\Delta E_{L}^{(NLO)} ( L , \{ \vec \phi _i = \vec 0 \} )
	= 
	\mathcal A ^{(NLO)}_{1, PB}
		\frac{ \exp \left( - \sqrt{\frac{8}{3}} \kappa L \right)}{ (\kappa L)^{3/2} } 
	+
	\mathcal A ^{(NLO)}_{2, PB}
		\frac{ \exp \left( - \sqrt{\frac{2}{3}} \kappa L \right)}{ (\kappa L)^{5/2} } 
	\ .
\end{equation}
As in the two-body case, we assume the amplitude to the NLO corrections to be of the size of the leading order amplitude and again set $\mathcal A ^{(LO)}_{ PB} {=} \mathcal A ^{(NLO)}_{i, PB}$ in the expressions above.
\par
As discussed in Section~\ref{sect:twists for N bodies}, the effects of general twisted boundary conditions in an $N$-body system is obtained by multiplying the wave function in a box by a phase related to the twists whenever one leaves the box. In the case of FV corrections in the three-body system, one can extend the result for periodic boundaries obtained by \cite{Meissner:2014dea} for general twists by including such a phase,
\begin{align}
	\Delta E_{L}^{(LO)} ( L ,\{ \vec \phi _i = \vec 0 \} )
	& \
	\mapsto
	\Delta E_{L}^{(LO)} ( L, \{ \vec \phi _i \})
	=
	\sum_{i=1}^3
	\sum_{   (\vec n _i , \vec n _j, \vec n _k) \in M_i }
		\hspace{-10pt } v_{PB} ( \vec n _i, \vec n _j , \vec n_k ) \ e^{ i \sum \vec \phi _n \cdot \vec n _n }\ .
\end{align}
Here the amplitude $v_{PB} ( \vec n _i, \vec n _j , \vec n_k )$ is associated with the interaction of the infinite volume bound state located in the original volume with its neighboring images,
\begin{equation}
	v ( \vec n _i, \vec n _j , \vec n_k )
	:=
	\intdd{\vec x _1} \intdd{\vec y _1} \psi( \vec x _1 , \vec x _2) V_i( x_i)
		\psi ( \vec x _i - \vec n _j + \vec n _k, \vec y _i + \frac{1}{\sqrt{ 3}} ( \vec n _j + \vec n _k - 2\vec n _i))\ .
\end{equation}
The vectors $\vec x _i$ and $ \vec y _i$ correspond to the three-body Jacobi coordinates where particle $i$ is isolated and the sets $M_i$ are defined such that vector tuples in this set minimize the hyper radius under certain constrains (e.g. one sets one of the vectors equal to a unit vector and the rest equal to zero),
\begin{equation}
	M_i
	:=
	\text{min} _\rho \left(
	\left\{
		(\vec n _i , \vec n _j, \vec n _k) \ \in \mathbb Z ^9
		\; \bigg| \;
		\vec n _j - \vec n_k \neq 0
	\right\}
	\right)
	\ .
\end{equation}
Thus, instead of a factor of $3 \times 2 \times 2 \times 3$ (spatial dimensions $\times$ sign of vector $\times$ permutations of $(j,k)$ for fixed $i$ $\times$ permutations of $i$) when executing the sum over all neighboring lattice vectors, one obtains a factor given by a sum of cosines depending on the twist angles\footnote{It is interesting to study this scenario in the $N$-body case as well. Since one expects no directional dependence of the leading order finite volume energy shift in the unitary limit of $N$-identical bosons, one can assume that the general twist dependence of this shift can be expressed by equation~\eqref{eq:analtic_amp3} when changing 3 to $N$. We are currently studying the twist dependence in the $N$-body fermionic case \cite{koerber2016}.},
\begin{align}\label{eqn:phi result}
	\Delta E_{L}^{(LO)}( L , \{ \vec \phi _i \} ) 
	&=
	\mathcal A ^{(LO)} (\{ \vec \phi _i \} )  \frac{ \exp \left( - \frac{2}{\sqrt{3}} \kappa L \right)}{ (\kappa L)^{3/2} } 
	\\ \label{eq:analtic_amp3}
	\mathcal A ^{(LO)} ( \{ \vec \phi _i \} )
	&= 
	\frac{ \mathcal A ^{(LO)}_{PB} }{9} \ \sum \limits_{i,j=1}^3 \cos ( \vec e _j \cdot \vec \phi _i )
	\ .
\end{align}

\begin{figure}[htb]
\centering
\includegraphics[width = .95\textwidth]{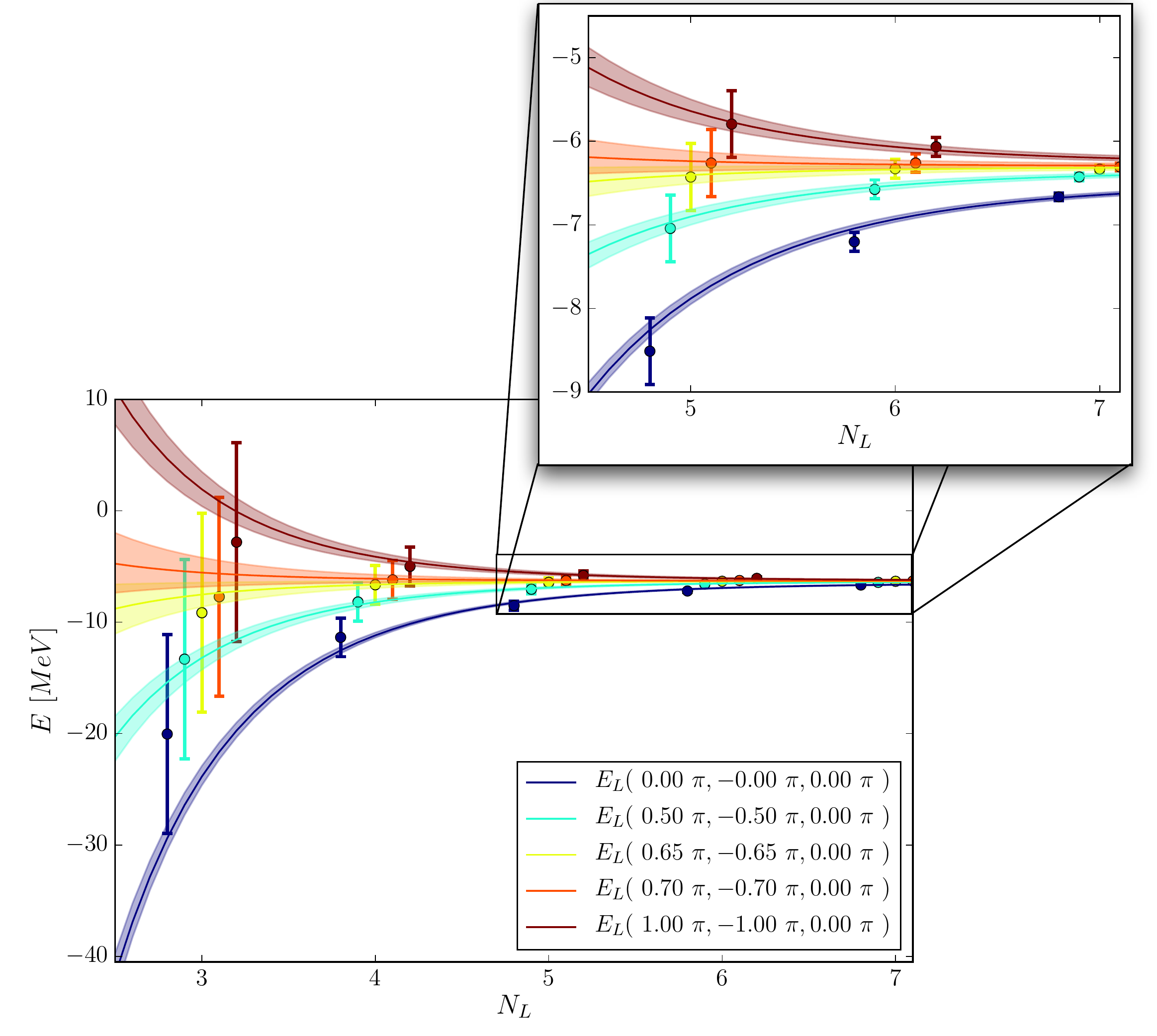}
\caption{\label{fig:triton_coefficients} Individually selected three-body fits for a fit range from $N_L \in[ 3, 7]$ which corresponds to $L = a N_L$ with $a = 1.97 $ fm. Extracted infinite volume energies, amplitudes and average $\chi^2$ can be extracted from figure \ref{fig:triton_results}. The error bars and error bands correspond to 1-$\sigma$. Data points and bands are slightly shifted in $N_L$ direction for visualization purposes.}
\end{figure}

\begin{figure}[htb]
\centering
\includegraphics[width = .95\textwidth]{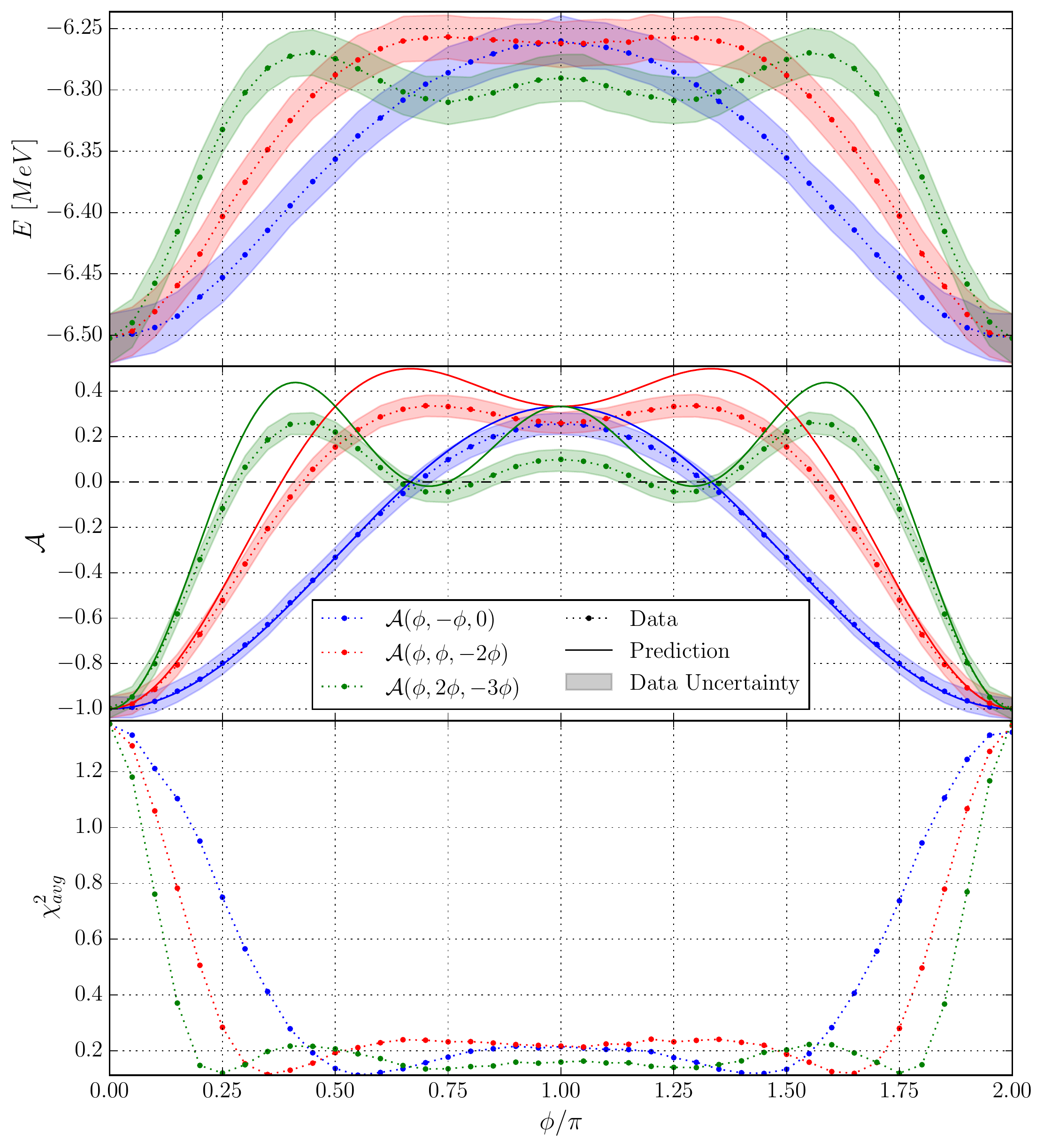}
\caption{\label{fig:triton_results}Triton results for individual twist fitting procedure. The figure contains the individually extracted infinite volume energy results (top), the amplitudes of the leading order FV behavior (center) as well as the average $\chi^2_{avg}$ (bottom) for each fit in a fitting range from $N_L \in [3,7]$. The data points are obtained for three different angle configurations (red, blue, green). Data points correspond to dots (and are connected via dotted lines), the uncertainties of the data points correspond to the bands and the solid lines are the predictions for the behavior of the amplitude. The error bands correspond to an 1-$\sigma$ confidence interval of the propagated errors.}
\end{figure}

Again, as in the two-body case, we restrict ourselves to equal boosts in each spatial direction $\vec e _j  \cdot \vec \phi _i = \phi _i \in [0, 2\pi]$ and the sum of all twist angles is constrained to zero to ensure zero CM motion. To analyze the twist-angle dependence of the leading order FV corrections, calculations were performed using three different angle orientations,
\begin{align}
	( \phi _1, \phi _2, \phi _3) =
	( \phi, -\phi, 0) \; ,\;
	( \phi,  \phi, -2 \phi) \; ,\;
	( \phi, 2\phi, -3\phi)
\end{align}
Selected energy fits from our calculated distributions are displayed in figure \ref{fig:triton_coefficients} and the corresponding amplitudes coming from all our fits as well as their predictions can be found in figure \ref{fig:triton_results}.

In contrast to the two-body case, the extrapolated infinite volume energy as well as their average $\chi_{avg}^2$ of individual fits depend on the twists, as can be seen by comparing the top and bottom panels of fig.~\ref{fig:triton_results}.  This is to be expected since we have used in our fits a FV functional form (equation~\eqref{eqn:meissner equation}) that does not represent our system exactly.  Errors in the fitted amplitudes and energies can, and most certainly are, correlated in this case (compare top and center panels of fig.~\ref{fig:triton_results}).  We can explicitly see how such correlations come about by considering the following example.  Let us assume that the exact leading order FV expression is parametrized by 
\begin{displaymath}
	\Delta E_{L, exact}^{(LO)}(L, \phi) 
	= 
	\mathcal{A}^{(LO)}(\phi)\left[ \Delta E_{L, used}^{(LO)}(L,0)+ \Delta E_{L, corr}^{(LO)}(L)\right]\ ,
\end{displaymath}
where $\Delta E_{L, used}^{(LO)}(L,0)$ is given by equation~\eqref{eqn:meissner equation} and $\Delta E_{L, corr}^{(LO)}(L)$ its correction.  The extracted infinite volume energy will have an explicit dependence on the twist angles since
\begin{align}
	\nonumber
	E( L, \phi )
		&= 
			E_\infty
			+
			\Delta E_{L, exact}^{(LO)}(L, \phi)
			+
			\delta E(L)
			\\ \label{eqn:error1}
		&= 
			E_\infty 
			+
			\mathcal{A}^{(LO)}(\phi)\left[ \Delta E_{L, used}^{(LO)}(L,0)+ \Delta E_{L, corr}^{(LO)}(L)\right]
			+
			\delta E(L)
			\\ \label{eqn:error}
		&=
			E_\infty +  \delta E_\infty(L, \phi)
			+
			\Delta E^{(LO)}_{L,used}(L, \phi)
			+
			\delta E(L)
			 \ .
\end{align}
The expression above shows how errors in the form of the fitting function $\Delta E_{L, used}^{(LO)}$ can induce twist angle dependence and correlations $\delta E_\infty(L, \phi)$ on our extrapolated energies $E_\infty$.  Comparing equations~\eqref{eqn:error1} and~\eqref{eqn:error}, we note that as $ \mathcal A ^{(LO)}(\phi)  \rightarrow 0$, one has $\delta E_\infty(L, \phi) \rightarrow 0$. Thus the fitting expression, regardless of form, becomes exact in this limit.

Returning to figure~\ref{fig:triton_coefficients}, we find that the shape of the leading order FV corrections is similar to that of the two-body deuteron system.  In particular, one finds energies converging from below and above the infinite-volume energy. However, because of the dimensionality of the problem, the accessible box sizes where not sufficiently large to enter the asymptotic region where the error bands overlap.

The amplitude fits in the center panel of figure~\ref{fig:triton_results} suggests that certain twist angle combinations have significantly reduced FV corrections (i.e. when $\mathcal{A}(\phi)=0$), similar to the iPBC case in the two-body system\footnote{The term `i-periodic' in the two-body case refers to angles that produce a purely imaginary phase (i.e. $\theta=\pi/2$), and that also significantly reduce LO FV effects.  For the three-body case, we designate the term `i-periodic analogues' to refer to twist angles that eliminate the leading order FV effects, but are not in general purely imaginary.}.  Indeed, from the predicted shape of the amplitude twist dependence in equations~\eqref{eqn:phi result} and~\eqref{eq:analtic_amp3}, the three-body analogue to iPBCs occurs for twist angles that solve the following equation,
		\begin{equation}\label{eq:twist_sol0}
			\sum \limits _{j=1}^3 \cos ( \vec \phi_1 \cdot \vec e _j ) 
										+  \cos ( \vec \phi_2 \cdot \vec e _j ) 
										+  \cos ( (\vec \phi_1 + \vec \phi_2) \cdot \vec e _j ) 
			= 0\ .
		\end{equation}
If one twists each spatial direction equivalently, this reduces to
		\begin{equation}
			\label{eq:twist_sol}
			\cos ( \phi_1  ) +  \cos (\phi_2) +  \cos ( \phi_1 + \phi_2 ) = 0\ .
		\end{equation}
Note that the analogue to iPBC twists are not unique in the three-body system, as opposed to the deuteron case.  In particular, there is a one-dimensional set of values for $\phi$ which eliminate the leading order FV effect.  In figure~\ref{fig:triton_results}, for example, the iPBC analogue twist angles correspond to the values of $\phi$ where the solid curves of $\mathcal{A}(\phi)=0$ (center panel). 

\begin{figure}[htb]
\centering
\includegraphics[width = .95\textwidth]{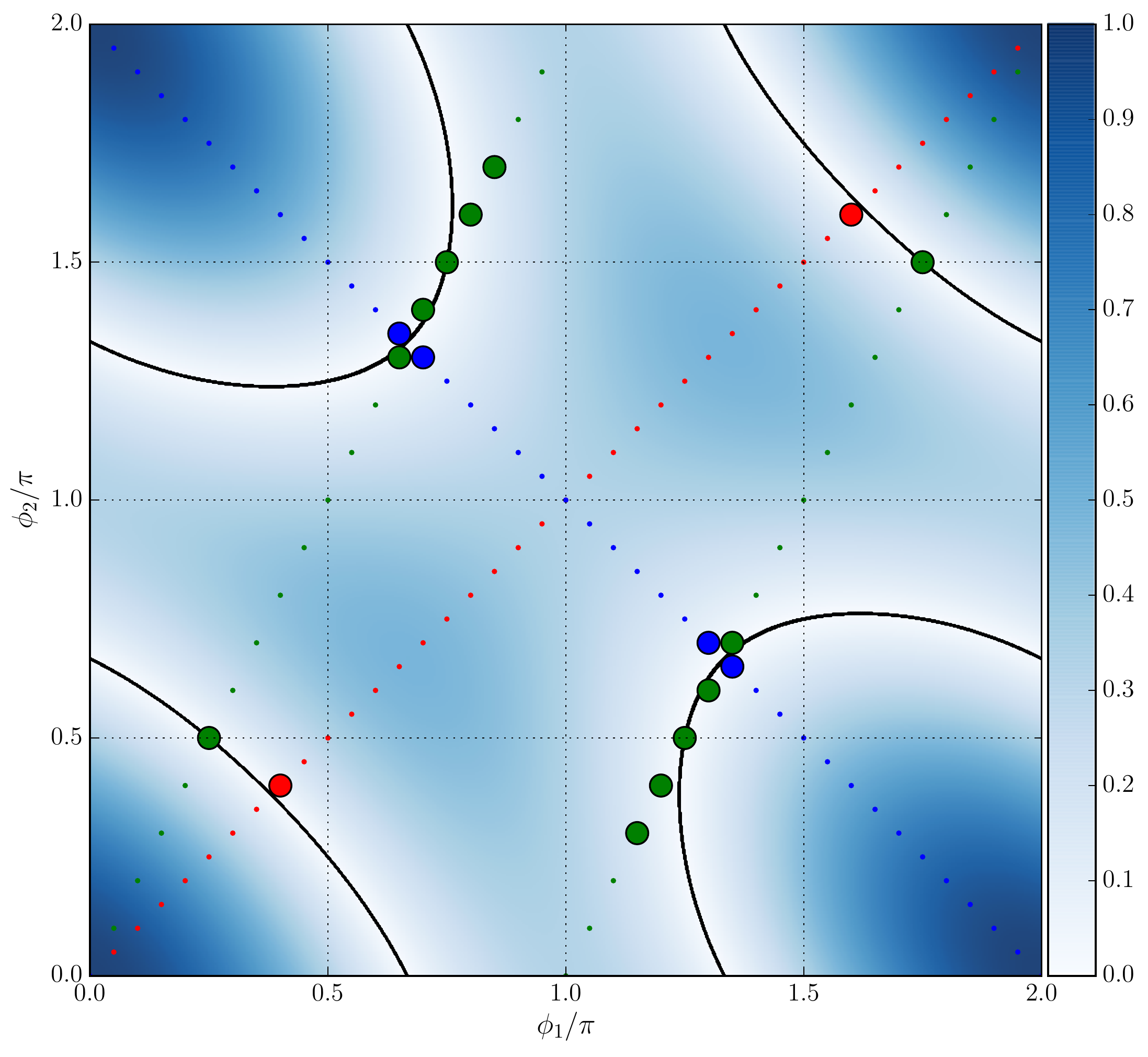}
\caption{\label{fig:ipbs3}Contour plot for three-body iPBC analogues. The contour expresses the relative amplitude of the predicted analytic LO FV corrections depending on the twists: $\mathcal A ^{(LO)} (\phi_1, \phi_2 , - \phi_1 - \phi_2) $ (equation~\eqref{eq:analtic_amp3}). The black line correspond to the solutions for the three-body iPBC analogues: $\mathcal A ^{(LO)} = 0$ (equation \eqref{eq:twist_sol}) and the points (both large and small) represent the twist angles we have used in our study. The larger points have bee identified with the numerically found iPBC analogues extracted from figure \ref{fig:triton_results} (see text). As in figure \ref{fig:triton_results} the color code represents data for a fixed ratio of $\phi_2 / \phi_1$: (blue, red, green) = ($-1$,$1$,$2$).}
\end{figure}

In figure~\ref{fig:ipbs3} we give a contour plot which shows all the allowed iPBC analogue twist values for the three-body system investigated in this paper. The contour itself represents the left-hand-side of equation~\eqref{eq:twist_sol} and thus the predicted relative amplitude of the LO FV corrections. The maximal values correspond to the dark blue regions (periodic and other boundaries) and the minima to the white regions (`i-periodic' boundaries). The solid line corresponds to the exact solutions of equation \eqref{eq:twist_sol} and the points are the locations of pairs of $(\phi_1, \phi_2)$ which were used for in our numerical investigations. The larger points give the numerically extracted twist angles that are consistent (within errors) with $\mathcal{A}(\phi)=0$, results which are valid to all orders of the FV corrections. Differences between our extracted points and the solid line are presumably attributed to higher order FV effects.  
The data points of figure~\ref{fig:ipbs3} use the same color designations as in figure \ref{fig:triton_results}.  The results in the middle panel of figure \ref{fig:triton_results} represent cross-sections of figure~\ref{fig:ipbs3} along the dotted lines and thus the data points share a  $2\pi$ periodicity.

Finally, we note that a complete averaging of twists does not completely remove FV effects in the three-body system, due to an offset in results that may be attributed to higher-order FV effects.  Because of this offset, we find that the results of iPBC analogues are superior to twist averaging for the three-body system.

\clearpage

\section{Conclusion\label{sect:conclusion}}

In this paper we investigated the effects of twisted boundary conditions on two- and three-body nuclear systems.  We performed investigations using the NLEFT formalism, with appropriate modifications to affect twisted boundaries and utilized a simplified NN interaction.  We benchmarked our two-body results to known analytic results from \cite{Briceno:2013hya}, and extended the analysis in the two-body sector to additional twist angles.  We performed the same analysis to the three-body (triton) sector, where we derived the three-body iPBC analogue quantization condition and numerically verified their corresponding FV cancellations.   As opposed to the deuteron case, we found multiple iPBC analogue twist possibilities in the three-body sector.      

We have attempted a detailed analysis of our fitting and extraction routines, where we enumerate all (known) sources of systematic errors.  Where possible, we assign realistic errors or very conservative errors in our error budget.  We find that, in both two-body and three-body systems, results obtained with iPBC analogues were superior to twist averaging, under the constraint that the allowed twists preserved zero CM motion.
In our three-body calculations, we found correlations and twist-angle dependence in our extrapolated binding energies.  As we demonstrated, this finding is to be expected since the leading order FV functional form we used to extract our results was derived for three particles at the unitary limit \cite{Meissner:2014dea}, which does not describe our system exactly.

Our work also shows that the implementation of twisted boundaries for $N$-body systems within the NLEFT formalism is, in principle, relatively easily done.  One simply has to multiply off-diagonal matrix elements of the transfer matrix by a phase associated with the twists. One might fear that this procedure increases sign oscillations during stochastic computations, however, the corresponding operators remain hermitian and the eigenvectors and corresponding eigenvalues remain real.  Furthermore, though in this case only two-body interactions were considered, this can be easily generalized to $N$-body interactions.

In this study we applied twists to nucleon degrees of freedom within a non-relativistic formulation. We found that the iPBC analogue twists exactly cancelled the leading order FV effects.  This is contrast to LQCD calculations that employ twists (or partial twists), since here the twists are applied directly to quarks.  The interactions in this case can also depend on the twist angles due to propagation of pions around the torus.  As such, we do not expect exact cancellation of leading order FV effects, but rather a suppression.  It would be interesting to quantify this level of suppression from LQCD studies that utilizes our iPBC analogue twist-angle condition in equations~\eqref{eq:twist_sol0} and~\eqref{eq:twist_sol}.

The most intriguing aspect of this work is the demonstration of iPBC analogue twists for the three-body sector, which have vastly reduced FV corrections compared to PBCs.  This raises the question of whether or not there exists more general iPBC analogue twists for $N$-body systems. The possibility of iPBC analogue twist angles for higher $N$-body systems would be an important finding for finite-volume numerical simulations since this would allow for calculations in smaller volumes accompanied by their significant reduction in computational costs.  Our findings in the three-body sector, coupled with our current investigations of more general $N$-body systems \cite{koerber2016}, provides credence that analogues of iPBC angles exist for non-relativistic $N$-body systems in general.

\section{Acknowledgments\label{sect:acknowledgments}}

We acknowledge financial support from the Magnus Ehrnrooth Foundation of the Finnish Society of Sciences and Letters, which enabled some of our numerical simulations.  We are indebted to S. K\"onig, T. L\"ahde and A. Shindler for insightful discussions. We thank D. Lee for initial discussions related to applying TBCs within the NLEFT formalism.

\appendix

\end{document}